\documentclass{pas}

\usepackage{amsmath}
\usepackage{multirow}
\RequirePackage{aas-macros}
\usepackage{orcidlink}
\usepackage[nopatch]{microtype}
\usepackage{booktabs}
\usepackage{listings}
\usepackage{graphicx}
\usepackage{graphbox} 
\usepackage{natbib}
\usepackage{hyperref}
\DeclareUnicodeCharacter{2212}{-}
\usepackage{lineno}
\usepackage{booktabs}
\usepackage{tabularx}
\usepackage{array}
\usepackage{yhmath}

\begin{document}
\lefttitle{Dot to dot}
\righttitle{Graham et al.}
\jnlPage{xxx}{xxx}
\jnlDoiYr{2025}
\doival{10.1017/pasa.xxxx.xx}

\articletitt{Research Paper}


\title{Dot to dot: high-$z$ little red dots in $M_{\rm bh}$--$M_{\rm \star}$ diagrams with galaxy-morphology-specific scaling relations}

\author{\sn{Alister W.} \gn{Graham}$^1$\orcidlink{0000-0002-6496-9414}, 
        \sn{Igor V.} \gn{Chilingarian}$^{2,3}$\orcidlink{0000-0002-7924-3253}, 
        \sn{Dieu D.} \gn{Nguyen}$^4$\orcidlink{0000-0002-5678-1008},   
        \sn{Roberto} \gn{Soria}$^{5,6}$\orcidlink{0000-0002-4622-796X}, 
        \sn{Mark} \gn{Durr\'e}$^1$\orcidlink{0000-0002-2126-3905}, and 
        \sn{Duncan A.} \gn{Forbes}$^1$\orcidlink{0000-0001-5590-5518}
        }

\affil{$^1$ Centre for Astrophysics and Supercomputing, Swinburne University of Technology, Hawthorn, VIC 3122, Australia.\\
$^2$ Smithsonian Astrophysical Observatory, 60 Garden St. MS09, Cambridge, MA 02138, USA.\\
$^3$ Sternberg Astronomical Institute, M.\ V.\ Lomonosov Moscow State University, 13 Universitetskiy prospect, 119234 Moscow, Russia.\\ 
$^4$Simons Astrophysics Group (SAGI) at International Centre for Interdisciplinary Science and Education (ICISE), Institute For Interdisciplinary Research in Science and Education (IFIRST), 07 Science Avenue, Ghenh Rang, 55121 Quy Nhon, Vietnam.\\
$^5$ INAF-Osservatorio Astrofisico di Torino, Strada Osservatorio 20, I-10025 Pino Torinese, Italy.\\
$^6$ Sydney Institute for Astronomy, School of Physics A28, The University of Sydney, Sydney, NSW 2006, Australia.}


\corresp{A.W.\ Graham, Email: AGraham@swin.edu.au}

\citeauth{Graham A.W., Chilingarian I.V., Nguyen D.D., Soria R., Durr\'e M., Forbes, D.A. (2024) Dot to dot: high-$z$ little red dots in $M_{\rm bh}$--$M_{\rm \star}$ diagrams with galaxy-morphology-specific scaling relations {\it Publications of the Astronomical Society of Australia} {\bf 00}, xxx--xxx. https://doi.org/10.1017/pasa.xxxx.xx}

\history{(Received 15 December 2024; revised xx xx xxxx; accepted xx xx xxxx)}



\begin{abstract}

The high redshift 'little red dots' (LRDs) detected with the {\it James Webb Space Telescope} are considered to be the cores of emerging galaxies that host active galactic nuclei (AGN). For the first time, we compare LRDs with local compact stellar systems and an array of galaxy-morphology-dependent stellar mass-black hole mass scaling relations in the $M_{\rm bh}$--$M_{\star}$ diagrams.
%
%
When considering the 2023-2024 masses for LRDs, 
they are not equivalent to nuclear star clusters (NSCs), with the latter having higher $M_{\rm bh}/M_{\star}$ ratios. However, the least massive LRDs exhibit similar $M_{\rm bh}$ and $M_{\rm \star,gal}$ values as ultracompact dwarf (UCD) galaxies, believed to be the cores of stripped/threshed galaxies. We show that the LRDs span the $M_{\rm bh}$--$M_{\rm \star,gal}$ diagram from UCD galaxies to primaeval lenticular galaxies. In contrast, local spiral galaxies and the subset of major-merger-built early-type galaxies define $M_{\rm bh}$--$M_{\star,gal}$ relations that are offset to higher stellar masses.  
Based on the emerging 2025 masses for LRDs, they may yet have similarities with NSCs, UCD galaxies, and green peas.
Irrespective of this developing situation, we additionally observe that low-redshift galaxies with AGN align with the quasi-quadratic or steeper black hole scaling relations defined by local disc galaxies with directly measured black hole masses. This highlights the benefits of considering a galaxy's morphology --- which reflects its accretion and merger history --- to understand the coevolution of galaxies and their black holes.  Future studies of spatially-resolved galaxies with secure masses at intermediate-to-high redshift hold the promise of detecting the emergence and evolution of the galaxy-morphology-dependent $M_{\rm bh}$--$M_{\star}$ relations.

\end{abstract}

\begin{keywords}
galaxies: bulges, 
galaxies: elliptical and lenticular, cD, 
galaxies: structure, 
galaxies: interactions, 
galaxies: evolution, 
(galaxies:) quasars: supermassive black holes
\end{keywords}

\maketitle

\section{Introduction}


After 3C~273 was determined to be receding at 1/6th the speed of light \citep[redshift $z=0.158$:][]{1963Natur.197.1040S}, Schmidt's quasars have been found at ever-increasing redshifts.  \citet{2021ApJ...914...36I} report on over 40 low- and high-luminosity quasars\footnote{Quasi-stellar radio sources (quasars) have strong radio emission, while QSOs do not.} and quasi-stellar objects (QSOs) at $z\gtrsim6$.  
While QSOs are predominantly blue, it was recognised that some dust-reddened active galactic nuclei (AGN) may have been overlooked in the past \citep[e.g.][]{1993ApJ...402..479F, 1995Natur.375..469W, 1996ApJ...468..556S} and very high numbers are now being found. 
In addition to the samples of possible and confirmed AGN at $5 < z < 9$ \citep[e.g.][]{2023ApJ...959...39H, 2024arXiv240610341A}, 
recent records include an AGN 
at $z=8.50$ \citep{2023ApJ...957L...7K}, 
$z=8.63$ \citep{2024arXiv241204983T}, 
the QSO CEERS\_1019 at $z=8.7$ \citep{2023ApJ...953L..29L}, and
GNz11 \citep{2023arXiv230801230M, 2024Natur.627...59M, 2023A&A...677A..88B}.
Due to their red rest-frame optical colour and compact appearance in {\it James Webb Space Telescope} ({\it JWST}) images\footnote{These nascent galaxies are sometimes spatially resolved \citep{2024arXiv241114383R}.  However, a comparison of LRDs with UCD galaxies in the size-mass diagram is left for a follow-up paper once more LRD host galaxy sizes become available.}, some of
these recently detected ($4 \lesssim z \lesssim 10$) red QSOs  have been dubbed `little red dots'
\citep[LRDs:][]{2023ApJ...954L...4K, 2023ApJ...957L...7K, 2024ApJ...963..128B, 2024arXiv240403576K, 2024ApJ...968...38K, 
  2024ApJ...963..129M, 2024ApJ...968....4P, 2024ApJ...974L..26Y}.\footnote{Some LRDs, but more likely globular clusters, may have started life as the `little blue dots' (LBDs) found by \citet{2017ApJ...851L..44E}.}

For years, it was reported 
\citep[e.g.][]{2011ApJ...742..107B, 2013ApJ...773...44W, 2015ApJ...799..164P,  
2016ApJ...816...37V, 2018ApJ...854...97D, 2020ApJ...888...37D,  
2020A&A...637A..84P} that high-$z$ galaxies with AGN reside above 
the original $z\sim 0$ near-linear 
supermassive black hole (SMBH) mass - host spheroid stellar mass
($M_{\rm bh}$-$M_{\star,sph}$) relation \citep{1998AJ....115.2285M}.
This was regarded as evidence that, over time, the galaxies' stellar populations must play
catch-up to the SMBHs for the systems to arrive at the $z\approx 0$ relation.
However, at least two factors were confounding the veracity of
whether or not any evolution with redshift had been detected.  The first was that the $z\sim 0$
$M_{\rm bh}$-$M_\star$ relation is not linear. For gas-rich systems, 
such as spiral (S) galaxies and (wet merger)-built lenticular
(S0) galaxies, 
the $M_{\rm bh}$-$M_\star$ relation is much steeper than linear
\citep{2012ApJ...746..113G, 2013ApJ...764..151G, 2013ApJ...768...76S}.
%
For the S galaxies, \citet{2016ApJ...817...21S} presented the steep `blue sequence'  
initially suggested by the data of \citet{2000MNRAS.317..488S} and subsequently better quantified by \citet{2018ApJ...869..113D, 2019ApJ...873...85D}. 
Selecting QSOs from massive gas-rich galaxies at high redshifts effectively samples the upper end of the steep non-linear relation and, therefore, samples from above the near-linear $z\sim 0$ relation.
The apparent higher $M_{\rm bh}/M_\star$ ratios measured
at higher redshift need not imply that there has been any
evolution.  This scenario 
can be appreciated by looking at \citet[][their Figure~13]{2021ApJ...914...36I}
and
\citet[][their Figure~9]{2023ApJ...954L...4K}. These authors correctly noticed that the past trend with redshift in the $M_{\rm bh}$-$M_\star$ diagram was due
to luminosity bias in past samples of high-$z$ AGN. The explanation stems from the quadratic or steeper $M_{\rm bh}$-$M_\star$ relation.

The second issue has been the need for more attention to galaxy components and morphology.  As stressed previously \citep[e.g.][]{2019ApJ...876..155S}, $M_{\rm bh}$-$M_\star$ scaling relations defined by grouping all galaxy types, or even just the early-type galaxies (ETGs), will be misleading.  The slope and zero-point calibration of such relations will depend on the random number of specific galaxy types in one's sample. For example, 
the reported relations for samples of `all ETGs' depend on the arbitrary fraction of primordial/primaeval\footnote{The term `primaeval' is used here to refer to the first type of galaxy to form prior to subsequent significant merger events that change the galaxy type. This is expected to be a disc galaxy due to the conservation of angular momentum in the contracting gas clouds \citep{1994ApJ...422...11E}, which need not but may be clumpy at high-$z$ \citep{Mowla2024}.} S0 galaxies \citep{Graham-triangal} versus wet-major-merger-built S0 galaxies, the number of ellicular (ES)\footnote{\citet{1966ApJ...146...28L} introduced the `ES' galaxy nomenclature for ETGs with intermediate-scale discs that do not dominate the light at large radii. \citet{2016ApJ...831..132G} introduced the name `ellicular'.} and elliptical (E) galaxies, and the number of brightest cluster galaxies (BCGs) built from multiple major mergers.
To fully address the topic of galaxy/black hole coevolution, one requires
knowledge of the galaxy-morphology-dependent $M_{\rm bh}$-$M_\star$ relations that have  progressively advanced over the last dozen years or so. 


There is an additional population of ultracompact dwarf (UCD) galaxies \citep[e.g.][]{1995ApJ...441..120H, 1999A&AS..134...75H, 2000PASA...17..227D, 2001ApJ...560..201P, 2010ApJ...722.1707M, 2011AJ....142..199B, 2011MNRAS.412.1627C, 2013MNRAS.433.1997P,   2020MNRAS.497..765F, 2020MNRAS.492.3263G} to consider.  
They are commonly thought to be the
remnant nuclei of threshed low-mass disc galaxies \citep{1988IAUS..126..603Z, 
  2001ApJ...552L.105B, 2003Natur.423..519D}, composed of the inner dense compact nuclear star cluster (NSC) 
from the progenitor galaxy and some residual galaxy stars
forming a larger secondary component.  
\citet{2024MNRAS.535..299G} suggested that LRDs might be somewhat akin
to UCD galaxies with NSCs.\footnote{AWG discussed LRDs overlapping (connecting?) with UCD galaxies (and NSCs) at the July 2024 conferences
\url{http://cosmicorigins.space/smbh-sexten}
and
\url{https://indico.ict.inaf.it/event/2784/}.
}
%
NSCs and the inner components of UCD galaxies
occupy a similar distribution in the $M_{\rm 
  bh}$--$M_{\rm \star,sph}$ diagram \citep{2020MNRAS.492.3263G}. They are, however, often excluded from $M_{\rm bh}$--$M_{\rm \star}$ scaling diagrams.  
This manuscript presents the central massive black hole mass versus the stellar mass of NSCs, the inner component of UCD galaxies, the bulge component of disc galaxies,  and the spheroidal component of E galaxies.  It additionally displays black hole mass versus the total stellar mass of NSCs and UCD galaxies, along with all types of galaxies with directly measured black hole masses. Several samples of local AGN have also been added.  Equipped with this background, we explore how LRDs compare.

Section~\ref{Sec_data} introduces the various data sets that appear in the $M_{\rm bh}$-$M_{\rm \star,sph}$ and $M_{\rm bh}$-$M_{\rm \star,gal}$ diagrams presented herein.  Nearby galaxies with directly measured black hole masses, including UCD galaxies (Section~\ref{Sec_sub1}), along with samples of low-$z$ AGN (Section~\ref{Sec_lowz}) and high-$z$ AGN, including LRDs (Section~\ref{Sec_sub3}), have representation. 
In Section~\ref{Sec_Disc}, we discuss the $M_{\rm bh}$-$M_{\rm \star}$ diagrams.  We start with a short recap of developments over the past decade (Section~\ref{Sec_nl}), breaking away from the original near-linear relations \citep[e.g.][and references therein] {1988ApJ...324..701D, 1989IAUS..134..217D, 1995ARA&A..33..581K, 2007MNRAS.379..711G} defined by predominantly ETGs. In addition to advances using relations not based on one-fit-for-all types of galaxy, which are skewed/biased by the random fractions of ETGs and late-type galaxies (LTGs, i.e.\ S galaxies), in one's sample, we discuss how not grouping the different ETGs has similar benefits for avoiding a biased relation and revealing valuable scientific information.
Section~\ref{Sec_LRDs_AGN} discusses the LRDs and their distribution in the $M_{\rm bh}$-$M_{\rm \star}$ diagrams. Finally, Section~\ref{Sec_toe} describes the location of lower-$z$ AGN in the diagrams, revealing that they follow the steep relations defined by local galaxies with predominantly inactive black holes.
A concise summary is given in Section~\ref{Sec_Sum}. 
Although almost all of the $z\approx0$ galaxies and star clusters have had their masses obtained from redshift-independent distances, 
the slight differences in cosmologies between the studies of more distant AGN are ignored given that this will not account for the broad trends and no quantitative analysis is performed here.

\section{Data}\label{Sec_data}


\subsection{(Predominantly) inactive stellar systems}

\subsubsection{Galaxies with directly measured black hole masses}\label{Sec_sub1}

A local ($z \sim 0$) sample of predominantly inactive galaxies with directly measured SMBH masses is described in a recent series of papers spanning \citet{Graham:Sahu:22a} to \citet{2024MNRAS.535..299G}.  Multicomponent
decompositions \citep{2016ApJS..222...10S, 
  2019ApJ...873...85D, 2019ApJ...876..155S} of galaxy images obtained by the {\it Spitzer Space Telescope} ({\it SST}) were performed, separating inner discs
\citep[e.g.][]{1998MNRAS.300..469S, 2007ApJ...665.1084B}  and bar-induced (X/peanut
shell)-shaped structures \citep[e.g.][]{1972MmRAS..77....1D,     
  2016MNRAS.459.1276C} from bulges.  Bars, rings, and ansae were modelled. 
In addition to the single exponential disc model, truncated and anti-truncated disc models were used, as were inclined disc models as required.  

The absence from the above sample of galaxies with (directly measured black hole masses and) $M_{\rm \star,sph} \lesssim 2 \times 10^{9}$ M$_\odot$ and 
$M_{\rm \star,gal} \lesssim 10^{10}$ M$_\odot$ is an observational bias due to difficulties measuring their central black hole mass.  However, galaxies with lower stellar masses of course exist
and are increasingly reported with directly measured SMBH masses.  These additional galaxies are individually named in the diagrams and displayed with a different (cyan) symbol because they were not used to derive the scaling relations shown there. 

The ETGs were separated into BCGs, E and ES galaxies 
\citep{Graham:Sahu:22b}, and S0 galaxies that were further separated into dust-poor and
dust-rich, which is a good indicator of their origin as either primaeval (no major
mergers, likely stripped of gas due to ram-pressure and thus `preserved' but with an ageing stellar population)\footnote{Given their coincident location in the $M_{\rm 
  bh}$--$M_{\rm \star,sph}$ diagram with S galaxies, a small handful of dust-poor S0 galaxies in the sample may be the gas-stripped and faded S galaxies proposed by \citet{1972ApJ...176....1G} and \citet{1973MNRAS.165..231D}.}  or built from a wet
major merger event likely involving an S galaxy \citep{Graham-S0, Graham-triangal}. 
These latter dust-rich systems, presumably with considerable neutral hydrogen gas content, are also big galaxies with massive black holes that do not remove their gas on short timescales.  For example, the dusty S0 galaxy Fornax~A is a 3-Gyr-old merger product with $\sim 10^9$ M$_\odot$ of H{\footnotesize I} gas \citep{2019A&A...628A.122S}.  However, not all major merger remnants have retained a dusty appearance.

In \citet{2024MNRAS.535..299G}, NGC~4697, NGC~3379, NGC~3091, and NGC~4649
were reclassified as S0 rather than E galaxies, with the adjustment supported
by kinematic maps of the galaxies.  Although not dust-rich, all but NGC~3091
--- the brightest galaxy in Hickson Compact Group No.\ 42 --- are recognised
mergers in prior work.  As detailed in \citet{2024MNRAS.535..299G}, their
anti-truncated discs are also likely a signature of their merger origin.
Two other S0 galaxies are regarded here as major-merger remnants, 
although they too are not classified \citep[see][]{Graham-S0} as dust-rich (i.e.\ dust=Y, that is, a strong yes).\footnote{The four dust `bins' (Y, y, n, N) were introduced and are described in \citet{Graham-S0}.} 
The first is NGC~5813\footnote{The potentially depleted stellar core in NGC~5813 \citep{2011MNRAS.415.2158R, 2014MNRAS.444.2700D} is thought to be formed from the merger of a binary black hole \citep{1980Natur.287..307B, 2004ApJ...613L..33G}. However, it may be worth remodelling this galaxy with an anti-truncated disc to check if the simpler S\'ersic bulge model will suffice once this is implemented.} (dust=y, widespread but weak) \citep{1995A&A...296..633H, 2015MNRAS.452....2K}. 
Although not (and probably more correctly "no longer") dust-rich, 
NGC~5813 is an old merger surrounded by a group-sized hot gas halo \citep{2015ApJ...805..112R}
that destroys dust, removes gas, and thereby inhibits star formation, preferentially impacting lower-mass galaxies.
The second is NGC~7457 (dust=N, that is, a strong no) with cylindrical rotation about its major axis, revealing that it, too,
experienced a merger, determined via other means to have occurred 2-3 Gyr ago \citep{2002ApJ...577..668S, 2008AJ....136..234C, 2011ApJ...738..113H, 2019MNRAS.488.1012M}. 
NGC~7457  displays an anti-truncated stellar disc and is somewhat unusual in that it is a merger that has lost its dust.  However, this is plausible given its relatively low galaxy stellar mass of around $10^{10}$ M$_\odot$ compared to $\sim10^{11}$ M$_\odot$ for the dust-rich merger-built S0 galaxies.
Finally, two ES,b galaxies (NGC~5845 and NGC~1332, with dust=n, that is, only nuclear dust)  
are {\it suspected} major merger remnants given their embedded intermediate-scale discs, as noted in Section~\ref{Sec_toe}.

While the following S0 galaxies are not considered to have been built by a major merger and are not dust-rich, they have, however, experienced a minor merger or accretion event, leaving them still largely `primaeval' in the sense of their stellar mass and structure, having not morphed into an S galaxy \citep{1966ApJ...146..810J, 2013ApJ...766...34D, Graham-triangal}.  They are  
NGC~2787, 
NGC~3998, and 
NGC~4026 
(all three categorised as dust=y), and 
NGC~1023, 
NGC~4762, and 
NGC~7332 
(all three with dust=N). 
References to works discussing the merger history of these galaxies are provided in \citet[][Table~2]{Graham-S0}. 

For those thinking there is a lot to keep up with, we do not disagree.  The notion that S0 galaxies are only faded or merged S galaxies \citep[e.g.\ see the discussion of the colour-mass diagram by][]{2014MNRAS.440..889S} has been supplanted with the recognition of an additional primaevel S0 population from which the S galaxies formed.  The high $M_{\rm bh}/M_{\rm \star,sph}$ ratios of these initial S0 galaxies, and their location in the $M_{\rm 
  bh}$--$M_{\star}$ diagrams, is what 
ruled out the previous two (faded or merged S galaxy) formation channels for this population of S0 galaxies.  This revelation has also resulted in re-drawing the evolutionary paths in the colour-mass diagram \citep{2024MNRAS.531..230G}.
How the LRDs relate with this explicitly identified primaeval population is explored, for the first time, in this work.

\subsubsection{UCD galaxies and NSCs}
\label{Sec_ucd_data}

A local sample of UCD galaxies and NSCs with directly measured SMBH masses has come
from \citet{2009MNRAS.397.2148G} and \citet[][and references therein]{2020MNRAS.492.3263G}.\footnote{The inner stellar component of M59-UCD3 may be a nuclear disc rather than an NSC.}
This sample includes the SMBH and NSC masses for the (stellar-stripped S0 and now) compact elliptical (cE) galaxy M32 \citep{2018ApJ...858..118N} and the flattened `peculiar'\footnote{NGC~205 has dust patches and young stars near its centre \citep{1998A&A...337L...1H, 2006ApJ...646..929M}.} dwarf ETG 
NGC~205 \citep{2019ApJ...872..104N}, plus an updated black hole mass for 
NGC~404 \citep{2020MNRAS.496.4061D}\footnote{This mass is questionable as estimates from reverberation mappings suggest a black hole mass that is an order of magnitude smaller \citep{2024MNRAS.530.3578G, 2024ApJ...976..116P}.} 
and NGC~4395 \citep{2019MNRAS.486..691B}.\footnote{The nuclear star cluster mass is taken from \citet{2015ApJ...809..101D}.}
The sample is supplemented with the NSCs in NGC~5102, NGC~5206 \citep{2018ApJ...858..118N, 2019ApJ...872..104N}\footnote{NGC~5102 and NGC~5206 would benefit from a bulge/disc decomposition rather than the single S\'ersic model that has been fit to the main galaxy. The nuclear star cluster, rather than the nuclear star cluster plus nuclear disc, is shown here.  As shown in \citet{2007ApJ...665.1084B}, both nuclear components are common in S0 galaxies \citep[e.g.][]{2013MNRAS.431.3364L}.} and NGC~3593 \citep{2022MNRAS.509.2920N}. 
UCD736 orbiting within the Virgo Galaxy Cluster \citep{2020ApJS..250...17L, 2025arXiv250300113T}
has also been added, as have the globular clusters B023-G078 around M31 \citep{2022ApJ...924...48P}
and $\omega$ Centauri \citep{2013MNRAS.429.1887D, 2024Natur.631..285H} for which the {\it Gaia}
 -Sausage/Enceladus host galaxy mass is used \citep{2023MNRAS.526.1209L}, as discussed by \citet{2024arXiv241111251L}. Finally, the dwarf spheroidal galaxy Leo I \citep{2008ApJ...675..201M} has been included, although it may not contain a massive black hole \citep{2024A&A...684L..19P}.

\begin{figure*}
\begin{center}
\includegraphics[trim=0.0cm 0cm 0.0cm 0cm, width=0.8\textwidth, angle=0]{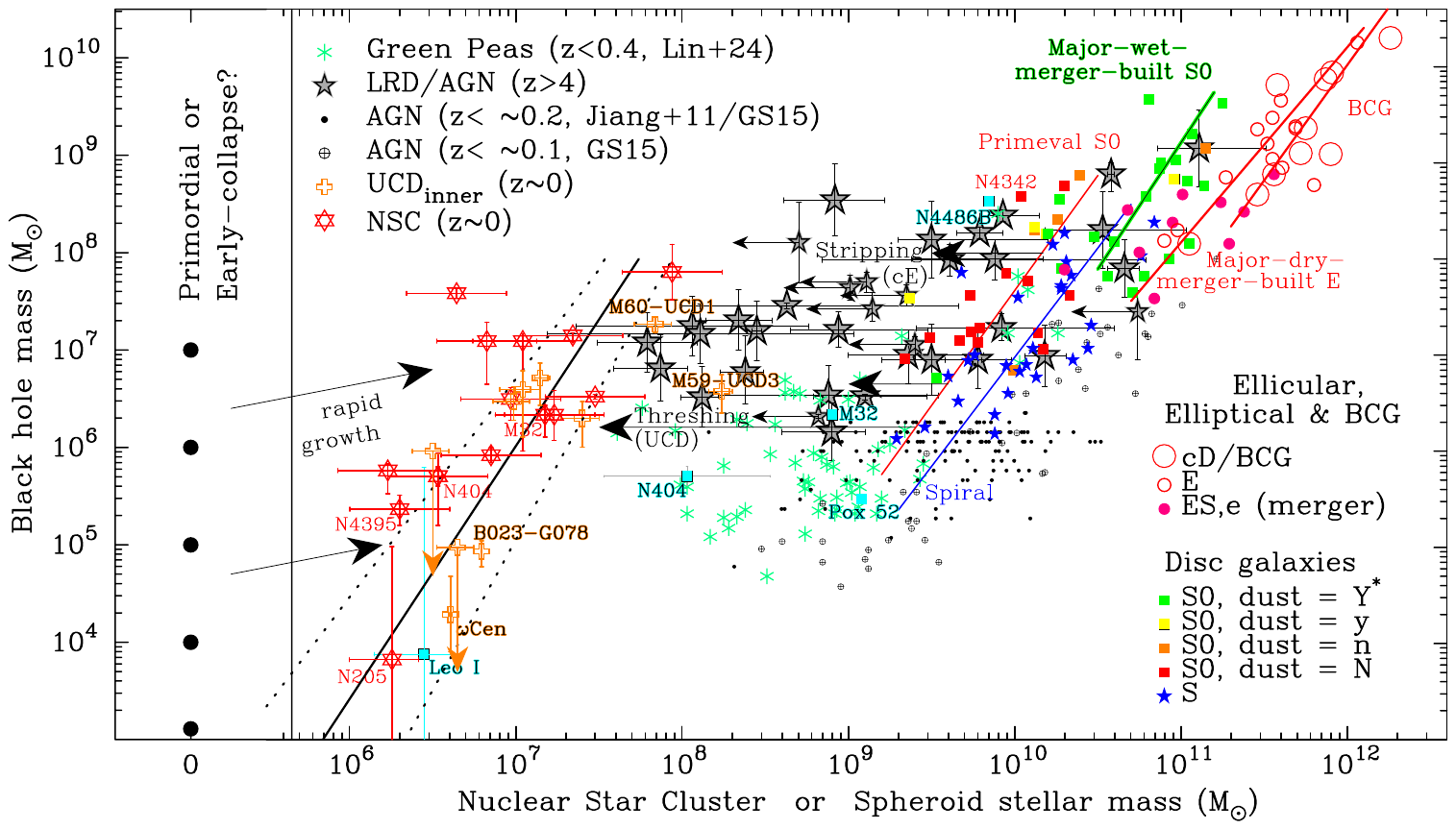}
\caption{$M_{\rm bh}$-$M_{\rm \star,sph}$ diagram and relations.
This is an extension of Figure~5 from \citet{2024MNRAS.535..299G}, itself an adaption of
Figure~6 from \citet{Graham:Sahu:22b}.  
From right to left, the lines from existing studies are as follows.
The right-most red line represents BCGs, and the second-from-right red line represents
non-BCG E galaxies \citep[][Table~2]{2024MNRAS.535..299G}, both primarily built from `dry' major mergers. 
The green line represents (`wet' major merger)-built dust-rich (dust=Y) S0 {\em and} the Es,b galaxies
\citep[][Table~2]{2024MNRAS.535..299G}, hence the asterisk on the Y in the figure legend. Next, 
the blue line represents S galaxies \citep[][Table~1]{Graham-triangal},
while the orange line represents dust-poor (dust=N) S0 galaxies referred to here as primaeval \citep[][Table~1]{Graham-triangal}.  Stripping 
and threshing the stars from these galaxies may produce cE and UCD galaxies, respectively. The
left-most solid and dotted lines represent the NSCs and inner component of UCD galaxies \citep[][Equation~6]{2020MNRAS.492.3263G}.
Notes: 
Spheroid masses of AGN with $4\times10^6 \lesssim M_{\rm bh}/M_\odot \lesssim 5\times10^7$
have likely been overestimated (in these suspected S galaxies, based on their
location in Figure~\ref{Fig-M-M-gal}). 
Without any structural decomposition of the LRDs, their total stellar mass is shown here under the implicit assumption, which we denounce (Section~\ref{Sec_LRDs_AGN}), that they are spheroidal structures without a disc component.
Upper-left legend:
Lin$+$24 = \citet{2024SCPMA..6709811L}; `LRD/AGN' covers the LRDs
reported in recent works, as noted in
Section~\ref{Sec_data}; Jiang$+$11 = \citet{2011ApJ...737L..45J}; GS15 =
\citet{2015ApJ...798...54G}.  The NSC and UCD data come from
\citet[][and references therein]{2020MNRAS.492.3263G}.
Lower-right legend: Galaxies with directly measured SMBH masses
\citep{Graham:Sahu:22a}, with updates noted in Section~\ref{Sec_data}.
Cyan squares (and the green peas and grey AGN samples) are additional galaxies not used to derive the relations.
}
\label{Fig-M-M-sph}
\end{center}
\end{figure*}

The original $M_{\rm bh}$--$M_{\rm \star,nsc}$ relation, involving nuclear star cluster masses, $M_{\rm \star,nsc}$, was first published in \citet{2016IAUS..312..269G}, having been presented at a 2014 IAU conference in Beijing. 
The relation was updated by \citet[][Equation~6]{2020MNRAS.492.3263G} and 
is shown in Figure~\ref{Fig-M-M-sph}, along with the galaxy-morphology-specific $M_{\rm bh}$--$M_{\rm \star,sph}$ relations. 
The former relation stems from the discovery of the $M_{\rm 
  \star,nsc}$--$M_{\rm \star,sph}$ relation \citep{2003ApJ...582L..79B,   
  2003AJ....125.2936G} coupled with the $M_{\rm bh}$--$M_{\rm \star,sph}$ relation.
  It is important to recognise that it appears to hold until
(i) the erosion of the star clusters at high black hole masses due to binary SMBHs
\citep{2010ApJ...714L.313B, 2012ApJ...744...74G} or
(ii) the appearance of a nuclear disc 10s-to-100s of parsec in size rather than
just an ellipsoidal star cluster.
To date,
the $M_{\rm bh}$--$M_{\rm \star,nsc}$ relation has not received anywhere near the
attention of the $M_{\rm bh}$--$M_{\rm \star,sph}$ relation, yet it holds
insight into the coevolution of dense star clusters and massive black holes
with important consequences for gravitational wave science from extreme
mass ratio inspiral (EMRI) events \citep[e.g.][]{2010ApJ...708L..42P, 
  2012A&A...542A.102M, 2017PhRvD..95j3012B, 2017JPhCS.840a2021G, 2023LRR....26....2A}.

For the $M_{\rm bh}$--$M_{\rm \star,sph}$ diagram shown in Figure~\ref{Fig-M-M-sph}, the stellar mass of the inner component of the UCD galaxies, i.e.\ the dense NSC obtained from decompositions of their light profiles, is displayed.  For the $M_{\rm bh}$--$M_{\rm \star,gal}$ diagram  (Figures~\ref{Fig-M-M-gal} and \ref{Fig-M-M-gal-plus}), the total stellar mass of the UCD galaxies is presented. 
%
%
This latter approach mirrors the plotting of the total stellar mass for the `ordinary' galaxies shown in Figures~\ref{Fig-M-M-gal} and \ref{Fig-M-M-gal-plus}, while their bulge or equivalently spheroid stellar mass is displayed in Figure~\ref{Fig-M-M-sph}.

\subsection{AGN with derived black hole masses}\label{Sec_sub2}

\subsubsection{Low-$z$ AGN}
\label{Sec_lowz}

\begin{figure*}
\begin{center}
\includegraphics[trim=0.0cm 0cm 0.0cm 0cm, width=0.8\textwidth, angle=0]{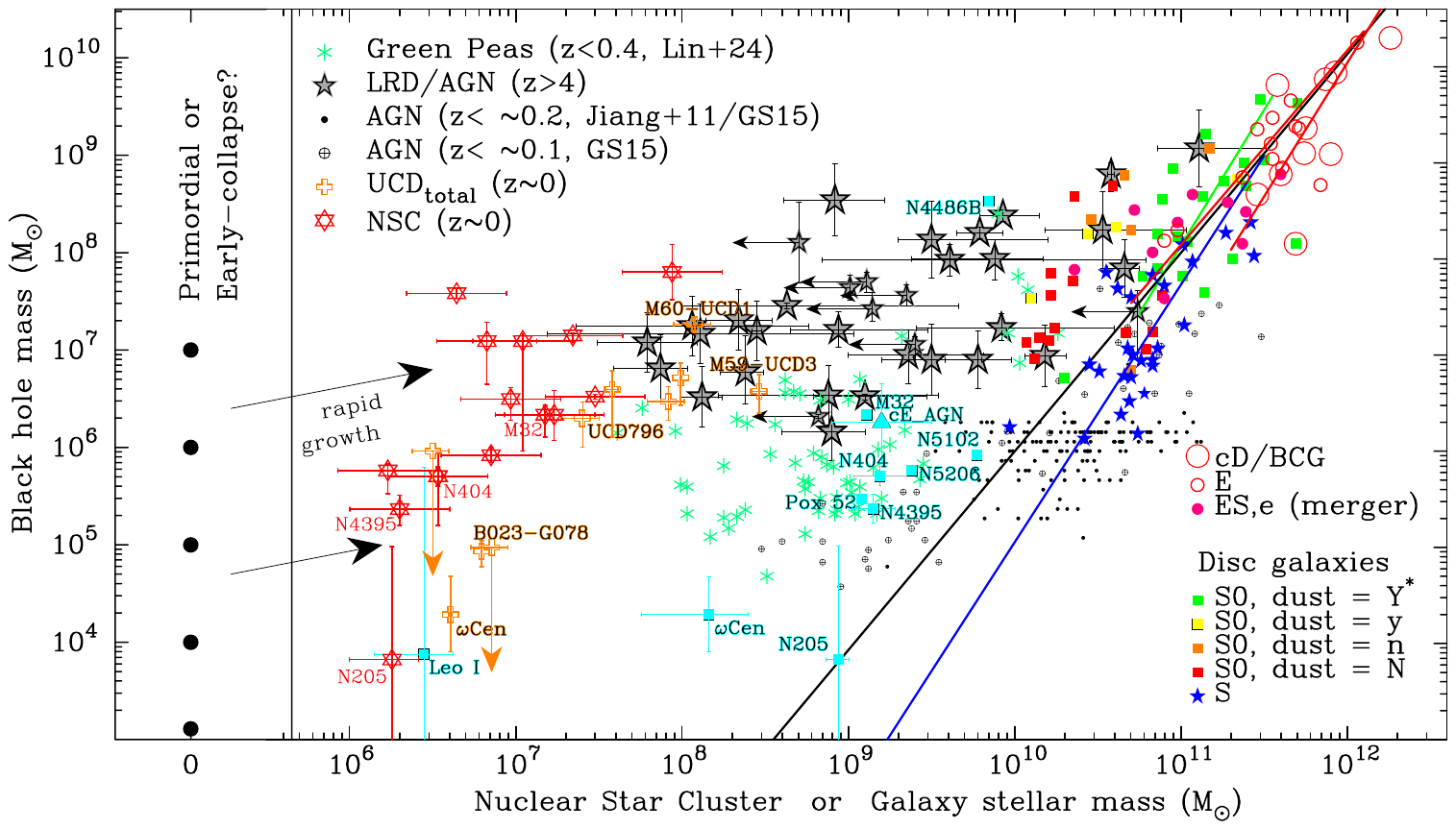}
\caption{$M_{\rm bh}$-$M_{\rm \star,gal}$ diagram and relations.
  Modification of Figure~\ref{Fig-M-M-sph}, building on Figure~A4 from \citet{Graham-triangal}.
  Here, the inner plus outer components of UCD galaxies are used for their galaxy stellar mass.
  The right-most red line (upper-right) denotes E BCGs, while the longer red line denotes
  non-BCG E galaxies \citep[][Table~2]{2024MNRAS.535..299G} and overlaps with
  the slightly steeper green line representing dust-rich (dust=Y) S0 and Es,b galaxies
\citep[][Table~2]{2024MNRAS.535..299G}.
  The long black line denotes galaxies built from major mergers
  \citep[][Table~1]{Graham-triangal}.
  The steepest (blue) line represents the S galaxies 
  \citep[][Table~1]{Graham-triangal}, which follow a steep `blue sequence'  \citep{2016ApJ...817...21S, 2018ApJ...869..113D}.  The very numerous (in the local Universe) dust-poor S0 galaxies (orange and red squares) with low stellar masses do not follow either of these relations.  The LRD/AGN sample shown here comes from the 2023-2024 data sets mentioned in Section~\ref{Sec_sub3}.}
  \label{Fig-M-M-gal}
\end{center}
\end{figure*}

Estimated black hole masses from the compilation of low-redshift ($z\lesssim0.2$)
AGN used by \citet{2015ApJ...798...54G} are presented in Figures~\ref{Fig-M-M-sph} and \ref{Fig-M-M-gal}.
The black hole mass estimates for the ten AGN from \citet{2013ApJ...775..116R} are
reduced here by 0.75 to bring the virial factor used by \citet{2013ApJ...775..116R}
in line with \citet{2011MNRAS.412.2211G}. The compilation also includes eleven AGN from
\citet{2014A&A...561A.140B}, ten AGN from \citet{2012ApJ...754..146M},
two from \citet{2014ApJ...782...55Y}, and UM~625 from \citet{2013ApJ...770....3J}.
This sample is bolstered with a further 8 (=10-2)\footnote{J153425.58$+$040806.7
and J160531.84$+$174826.1 \citep[from][]{2018ApJ...863....1C} are already included in the sample from 
\citet{2013ApJ...775..116R}.}
confirmed AGN with $0.024<z<0.072$ from \citet[][their Table~2]{2018ApJ...863....1C}.
AGN data from \citet{2011ApJ...737L..45J}, with $2\times 10^5 \lesssim M_{\rm 
  bh}/M_{\odot} \lesssim 2\times 10^6$, are also included, 
as is the cE galaxy SDSS~J085431.18+173730.5 \citep{2016ApJ...820L..19P} in Figure~\ref{Fig-M-M-gal}.

There are 1-sigma uncertainties in individual AGN masses of a factor of 2 to 3 due to estimates based on Doppler-broadened Gaussian emission lines \citep{2004ApJ...615..645O}.
The stellar mass-to-light ($M/L$) ratios are subject to systematic uncertainties related to the shape of the stellar initial mass function (IMF), internal dust attenuation, and mean stellar ages or star formation histories.
The metallicity affects the amount of gas cooling (from emission lines) and in turn impacts the fragmentation of gas clouds into stars and, thus, the IMF.\footnote{In a relatively metal-poor early-Universe, the IMF is top-heavy, where as in today's relatively metal-rich environments, the IMF is bottom heavy.} 
Collectively, this can yield a factor of 2 uncertainty on the adopted stellar masses.  Furthermore, 
the mass of the evolved stellar populations in today's galaxies may have been greater before the stellar-to-gas mass conversion arising from stellar winds and supernovae during the galaxy ageing process. Of course, some fraction of this second-generation gas may have turned into new stars.  To avoid crowding, these uncertainties are not shown in the figures. 

In passing, it is noted that \citet{2013ApJ...770....3J} report that the
galaxy UM~625 with an AGN can be convincingly categorised as hosting a
pseudobulge due to its blue colour and S\'ersic index $n<2$. However,
`classical' bulges built from major wet mergers will have lingering star
formation \citep{2024MNRAS.52710059G}, and their bulges can have $n<2$
\citep[e.g.][]{2020ApJ...903...97S, 2024MNRAS.535..299G}. 
Moreover, the high bulge-to-total ($B/T$) ratio of 0.6 reported by
\citet{2013ApJ...770....3J} additionally favours a merger origin.  UM~625 likely
represents an endemic misclassification of bulges, an issue raised by \citet{2014ASPC..480..185G}.  This situation is reminiscent of that with other dust-rich S0 galaxies built from major wet mergers, such
as NGC~1194, NGC~1316, NGC~5018 and NGC~5128, which reside below the old
near-linear $M_{\rm bh}$--$M_{\rm \star,sph}$ relation for ETGs but on the steep 
$M_{\rm bh}$--$M_{\rm 
  \star,sph}$ relation for merger-built (dust-rich) S0 galaxies \citep{Graham-S0}.

It is additionally noted that {\em galaxy} stellar masses were used for some
of the 34 AGN with $z\sim0.1$ in the $M_{\rm bh}$--$M_{\rm \star,sph}$ diagram
(Figure~\ref{Fig-M-M-sph}).  Specifically, the galaxies from
\citet{2014ApJ...782...55Y} and \citet{2013ApJ...775..116R}
were, in the absence of bulge/disc decompositions, 
effectively taken to be E galaxies. If they are S or S0 
galaxies, then the above practice will have acted to shift them rightward in Figure~\ref{Fig-M-M-sph}. 
Such a shift also occurs when using bulge masses if the
$B/T$ ratios of the disc galaxies have been overestimated.  This appears to be
the situation with the AGN data from \citet{2014A&A...561A.140B} and
\citet{2012ApJ...754..146M} based on the agreement of their AGN sample with
the inactive S galaxies in the $M_{\rm bh}$--$M_{\rm \star,gal}$ diagram
(Figure~\ref{Fig-M-M-gal}).

\begin{figure*}
\begin{center}
\includegraphics[trim=0.0cm 0cm 0.0cm 0cm, width=0.8\textwidth, angle=0]{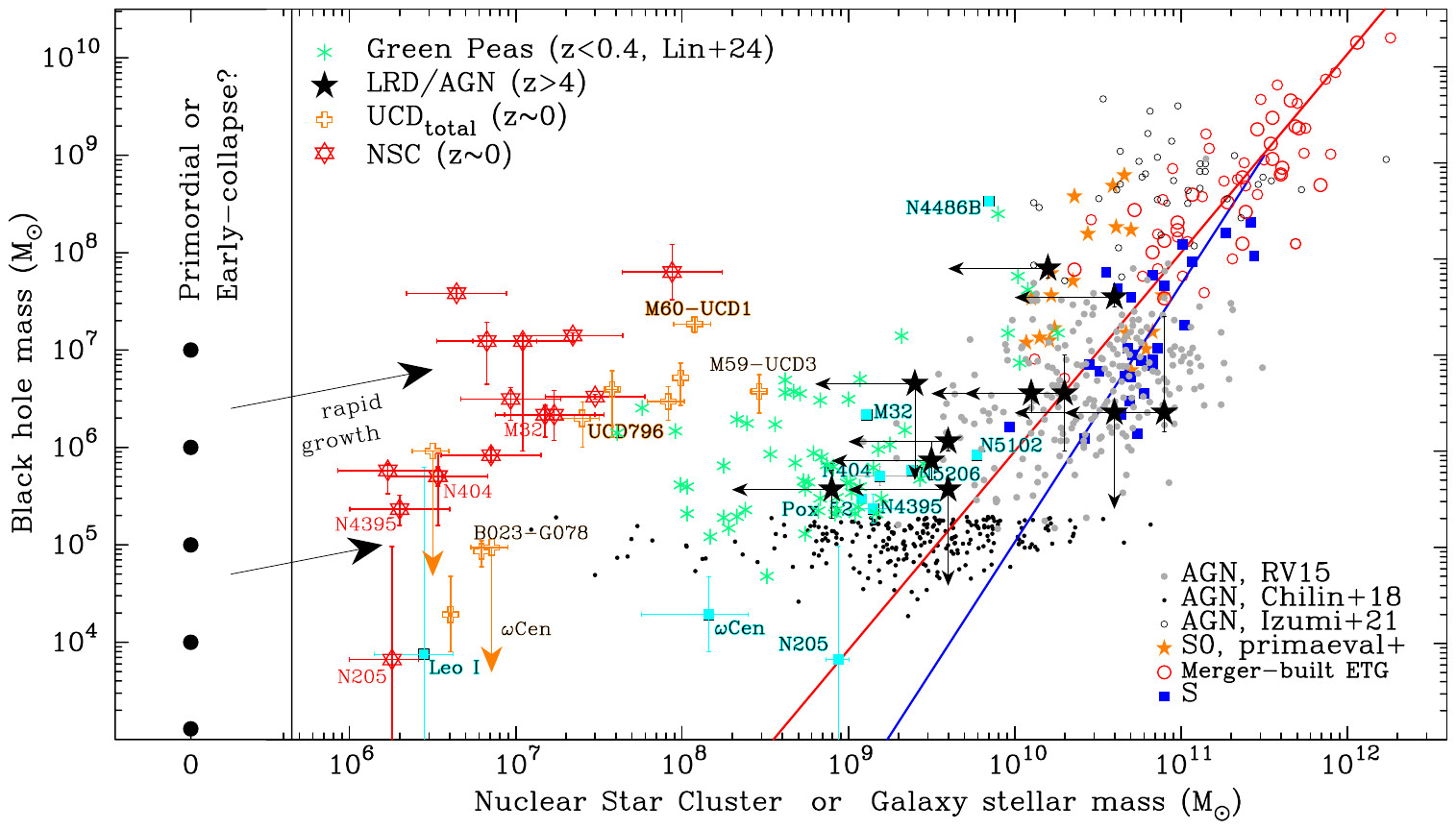}
\caption{Modification of Figure~\ref{Fig-M-M-gal}.
The quasi-quadratic (black) line represents galaxies built from major mergers
  \citep[][Table~1]{Graham-triangal} while the quasi-cubic (blue) line represents S galaxies 
  \citep[][Table~1]{Graham-triangal}, known to follow the steeper `blue sequence' discovered by \citet{2016ApJ...817...21S}. 
The $z<0.055$ AGN (larger
grey dots) from \citet{2015ApJ...813...82R} have been added; they 
support the steep quadratic/cubic $M_{\rm bh}$-$M_{\rm \star}$
relations; that is, they are not an offset population. 
The smaller black dots are low-$z$ galaxies with AGN hosting suspected intermediate-mass black holes \citep{2018ApJ...863....1C}.
The open black circles are $z\gtrsim6$ AGN with dynamical, rather than stellar,
galaxy masses \citep{2021ApJ...914...36I}. 
Lower-right legend: RV15 = \citet{2015ApJ...813...82R}; Chilin'$+$18 =
\citet{2018ApJ...863....1C}; Izumi$+$21 = \citet{2021ApJ...914...36I};
S = spiral galaxies with a directly-measured SMBH mass.
`Merger-built ETG' = S0, ES, E and BCG with a directly-measured SMBH mass and
known to have been built from a major merger;
`S0, primaeval$+$' = dust-poor low-mass galaxies with a directly-measured SMBH mass
and not known to have experienced a major merger; the $+$ acknowledges that some  
of these overlap with the distribution of S galaxies and as such are likely to be faded S galaxies rather than faded/preserved S0 galaxies that never sufficiently grew to host a spiral pattern. The LRD/AGN sample shown here is from \citet{2025arXiv250316595R} and has only upper limits for the stellar masses.
}
\label{Fig-M-M-gal-plus}
\end{center}
\end{figure*}

A second $M_{\rm bh}$--$M_{\rm \star,gal}$ diagram
(Figure~\ref{Fig-M-M-gal-plus}) has been made with alternate sources of AGN data.  This was done, in part, to avoid crowding.  The other (main) reason was to provide a diagram that better distinguishes the primaeval S0, S, and merger-built galaxies with directly measured black hole masses. The
estimated black hole masses in the AGN from \citet{2015ApJ...813...82R} and
the entire sample of 305 intermediate-mass black holes candidates 
with $4\times 10^4 \lesssim M_{\rm bh}/M_{\odot} \lesssim 2\times 10^5$ from
\citet{2018ApJ...863....1C} are shown in Figure~\ref{Fig-M-M-gal-plus}.\footnote{Aside from the ten ($8+2$) AGN from \citet{2018ApJ...863....1C} that are mentioned above and shown in Figure~\ref{Fig-M-M-gal}, just one other AGN (SDSS~J122548.86$+$333248.7) from \citet{2015ApJ...798...54G} appears in the larger sample from \citet{2018ApJ...863....1C} presented in Figure~\ref{Fig-M-M-gal-plus}.}  The SMBH masses from \citet{2015ApJ...813...82R} are reduced here by 0.7 (=0.75/1.075) to ensure a consistent virial factor, $f=3$, with the above AGN samples.
In addition, data from \citet{2024SCPMA..6709811L}  
for 59 `green peas'\footnote{`Green peas' are luminous but low-mass ($\lesssim 10^{10}$ M$_\odot$) compact galaxies with substantial star formation \citep{2009MNRAS.399.1191C}.} with $z < 0.4$ are shown, as are the $z\gtrsim 6$ AGN data
from \citet{2021ApJ...914...36I}, as noted in the following subsection.


\subsubsection{High-$z$ AGN and LRDs}\label{Sec_sub3}

Many teams have reported measurements of the stellar mass associated with high-$z$ AGN.  Before {\it JWST} data, \citet{2018PASJ...70...36I} and \citet{2021ApJ...914...36I} presented a compilation of AGN at $6 \lesssim z \lesssim 7$.  They stressed how sample selection bias of bright QSOs at these high redshifts had
swayed the conclusions from past investigations.  Their compilation rectified this situation by including lower luminosity QSOs at the same high redshifts.  Their data compilation is shown
here in Figure~\ref{Fig-M-M-gal-plus}, although it is noted that the galaxy
masses are dynamical rather than stellar, and, as such, smaller symbols have been
used for this sample.
Their SMBH masses were based on the prescriptions given by
\citet{2006ApJ...641..689V},  calibrated to a virial factor $f=5.5$
\citep{2004ApJ...615..645O}.
Here, these SMBH masses have been reduced by a factor of 3/5.5, which is in agreement with the calibration used by \citet{2015ApJ...798...54G}.

{\it JWST} has enabled the detection of AGN in LRDs over a wide range of high redshifts ($\sim$3-4 to $\sim$9-11).
Image resolution inhibits our ability to discern multiple components in these LRDs.  
LRD data from the studies listed below have been included in Figures
\ref{Fig-M-M-sph}--\ref{Fig-M-M-gal}, with the {\it same} total stellar mass presented in both diagrams.  The associated error bars are taken from the published works. 

The high-$z$ galaxy data shown here are from
\citet[][one AGN at $z=8.502$]{2023ApJ...957L...7K},  
\citet[][one AGN at $z=7.045$]{2024Natur.628...57F},  
\citet[][one AGN at $z=8.679$]{2023ApJ...953L..29L},  
\citet[][one AGN at $z=8.632$]{2024arXiv241204983T},  
\citet[][GNZ11 at $z=10.603$ = GNZ11]{2024Natur.627...59M},  
\citet[][one AGN at $z=4.658$]{2023Natur.619..716C},  
\citet[][one AGN at $z=5.55$]{2023A&A...677A.145U},   
\citet[][one dormant black hole at $z=6.68$]{2024Natur.636..594J},  
\citet[][two AGN at $z=6.34$ and 6.40]{2023Natur.621...51D},   
\citet[][two AGN at $z=5.242$ and 5.624]{2023ApJ...954L...4K},   
\citet[][two AGN at $z=6.68$ and 6.98]{2024ApJ...969L..13W}\footnote{We took the `medium' stellar masses and associated AGN masses from \citet{2024ApJ...969L..13W}, noting the data in their Table~1 does not match that in their Figure~8}, 
\citet[][9 AGN at $4<z<6$ plus one at $z=6.936$]{2023ApJ...959...39H},   
and 
\citet[][12 new AGN at $4<z<7$]{2023arXiv230801230M}.   


Uncertainties in the wavelength-dependent contributions of AGN and starlight to the spectral energy distribution (SED) can influence the $H_\alpha$,  $H_\beta$, and He{\footnotesize I} broad line equivalent width measurements and thus impact the AGN masses \citep[e.g.][]{2024ApJ...969L..13W}.  Moreover, after submitting this paper, the published stellar and black hole masses of these high-$z$ AGN continued to be debated as researchers probed the SED and the origin of the line broadening \citep{2025arXiv250316595R, 2025arXiv250316600D, 2025arXiv250316596N}. 
As with NGC~1068 \citep{1991ApJ...378...47M} and NGC~4395 \citep{2006ApJ...643..112L}, \citet{2025arXiv250316595R} report that the primary line-broadening mechanism in the distant AGN/LRDs appears to be electron scattering through a dense Compton-thick ionised gas (producing lines with exponential profiles) rather than Doppler motions (producing Gaussian or centrally broad profiles) that were found to play a lesser role. They show how this reduces the estimated black hole masses by 2 orders of magnitude, although it remains unclear what the revised stellar masses are, with \citet{2025arXiv250316595R} only reporting massive upper limits.  The actual stellar masses are expected to be smaller because not enough time has elapsed for such massive galaxies to arise.  Their data is shown in Figure~\ref{Fig-M-M-gal-plus}.  In passing, it is remarked that some of the high-$z$ quasars and QSOs reported to have very large black holes over the past 10-20 years will also have had their black hole masses overestimated {\em if} the broadening of their emission line(S) is mostly due to electron scattering. It may, therefore, be worthwhile reexamining the shape of the line profiles and checking for a broad plus narrow component in some of those systems.

\section{Context Setting and Discussion of Results}\label{Sec_Disc}

\subsection{Background Briefing}\label{Sec_nl}

For those new to the evolving field of black hole scaling relations, a recap of select relevant developments 
over the past decade may be helpful.  Other readers may wish to jump to section~\ref{Sec_LRDs_AGN}.

As explained in \citet{2012ApJ...746..113G}, 
it is understood why a near-linear $M_{\rm bh}$--$M_{\rm \star,sph}$ 
relation is inadequate to describe `ordinary' 
galaxies with `classical' (merger built) bulges \citep[as observed by][]{1998ApJ...505L..83L, 2001ApJ...553..677L, 1999ApJ...519L..39W, 2000MNRAS.317..488S}.  It is now recognised that each galaxy type follows a notably steeper than linear distribution \citep{Graham-triangal}, and 
recognition of galaxy-morphology-specific relations has enabled 
considerable breakthroughs in our understanding of galaxy evolution. For 
example, recognition of the galaxy-morphology-dependent $M_{\rm bh}$--$M_{\rm 
  \star,sph}$ relations has led to the identification of two types of  
S0 galaxies (those known to have low-metallicities and old ages are referred to as primaeval
\citep{2003AJ....125...66C, 2006AJ....132.2432L, 2013MSAIS..25...93S};  
the other is built from wet major mergers \citep{Graham-S0}).  In the past, these primaeval galaxies may have accreted gas and experienced minor mergers in such a way that they morphed into S galaxies, or a substantial collision may have led them to bypass such a transition and become a second-generation major-merger-built S0 galaxy, which can also be made from S galaxy collisions.  
\citet{Graham-triangal} has suggested that these low-mass, dust-free S0 galaxies are 
preserved, albeit faded, primaeval galaxies.  Ram-pressure stripping of cold gas from these galaxies within the hot X-ray emitting gas clouds/haloes of galaxy groups and clusters can act to preserve them by shutting down star formation.   The high velocities of the galaxies in the clusters inhibit  mergers of the low-mass galaxies and prevent this avenue of evolution. The cluster environment effectively `pickles' these galaxies.\footnote{Some distant LRDs have been observed in overdense regions \citep[e.g.][]{2022ApJ...930..104L, 2024arXiv241111534S, 2025arXiv250117925M}.} 
The
bulk of this galaxy type, often referred to as dwarf ETG
(dETG), are known to be nucleated \citep[e.g.][]{1985AJ.....90.1681B, 1985AJ.....90.1759S}, 
with even more found using the {\it Hubble Space Telescope} \citep[e.g.][]{2001ApJ...552..572L, 2003AJ....125.2936G, 2006ApJS..165...57C}.  As galaxy mass functions show, they, and dwarf irregular galaxies, are the most abundant galaxy type in the Universe today \citep[e.g.][]{1983ApJS...53..375R, 1987MNRAS.229..505P}. 
If they contain massive black holes, mergers of these systems, likely in the pre-cluster field and group environment, erode the central star cluster and thereby reduce the central surface brightness \citep{2010ApJ...714L.313B}.

As early as \citet{1977BAAS....9..347R}, some of these dETGs were considered dS0 disc galaxies. 
This has contributed to replacing the 
Tuning Fork diagram with an evolutionary scheme, dubbed the `Triangal', linking the galaxy types through accretions and mergers \citep{Graham-triangal}. It is the objective of this manuscript to begin to place LRDs in this greater context of galaxy evolution. 
The impact of mergers is widespread and wide-scale, from the erosion of NSCs 
to a potential `merger bias'\footnote{`Merger bias' is a term introduced here to capture how luminous red galaxies (LRGs) merge and thus reduce in number over time when building BCGs in clusters that trace the BAO boundary walls. At the same time, new LRGs, such as dust-rich S0 galaxies, form from lower-mass mergers in groups (with lower velocity dispersions congenial to mergers) that are initially out of clusters but fall towards them over time \citep{2011MNRAS.416.3033S, 2014MNRAS.442.2131A, 2015JKAS...48..213K}. Such spatial evolution in the distribution of LRGs due to mergers might introduce an evolving bias with redshift that skews BAO `standard ruler' measurements, complicating claims about an evolving dark energy equation of state.} affecting measurements of the baryonic acoustic oscillations (BAO) used to probe cosmology \citep{2005ApJ...633..560E, 2007MNRAS.381.1053P}.\footnote{The abundance of AGN with redshift offers another probe of the time evolution of the equation of state parameter \citep{2012MNRAS.420.2429L}.}  As we shall see in Section~\ref{Sec_toe}, these morphology-specific relations are also important for understanding the distribution of AGN in the $M_{\rm bh}$--$M_{\rm \star}$ diagram.  The distribution of the $z=0$ primaeval S0 galaxies is additionally important for checking potential connections with the distant AGN/LRDs, as done in Section~\ref{Sec_LRDs_AGN}.

Finally, it may be helpful to note that a detected offset in an $M_{\rm \star}$-$\sigma$ diagram between ETGs with and without directly measured black hole masses, which had potentially implied a bias in the $M_{\rm bh}$--$M_{\rm \star}$ relations derived from galaxies with directly measured black hole masses \citep{2016MNRAS.460.3119S}, was entirely due to a mismatch in the stellar light-to-mass conversion between the data sets \citep{2023MNRAS.518.1352S}. The $M_{\rm bh}$--$M_{\rm \star}$ relations should be applied as is to galaxies without directly-measured black hole masses; one simply needs to ensure that the stellar masses of one's sample are derived consistently with that used to establish the scaling relation.\footnote{The $M_{\rm bh}$--$M_{\rm \star}$ relations for the galaxies shown here were obtained using a diet-Salpeter IMF \citep{2001ApJ...550..212B}. Conversion factors for other IMFs are available in \citet{2024MNRAS.530.3429G}.}

\subsubsection{Compact massive red nuggets}

Although not highlighted in the figures, we mention the 
local massive `compact galaxies' \citep{1968cgcg.bookR....Z, 1971cscg.book.....Z} 
given that some works \citep[e.g.][]{2014ApJ...780L..20T, 2015A&A...578A.134S} 
are renaming these `relic galaxies' due to their similarity to the 
compact massive galaxies at
$z\sim2.5\pm1$, aka `red nuggets' \citep{2005ApJ...626..680D, 2008ApJ...677L...5V, 2011ApJ...739L..44D}.
They have sizes and masses comparable to the bulges of today's merger-built dust-rich S0 galaxies
\citep{2013pss6.book...91G, 2023MNRAS.519.4651H} and encapsulate the ES,b galaxies shown in \citet{Graham:Sahu:22b}.
Some red nuggets were found at $z<0.6$ using {\it SDSS} data \citep{2014ApJ...793...39D}, with one of the objects being an extremely compact post-starburst galaxy at $z=0.5$. 
Here, the four ES,b galaxies (with {\it Spitzer Space Telescope} imaging and multicomponent decompositions) 
have been grouped with the dust-rich S0 galaxies.  

For readers unfamiliar with the ES galaxy type introduced by (Martha)
\citet{1966ApJ...146...28L}, they have embedded, intermediate-scale stellar discs that
do not dominate the light at large radii \citep[e.g.][]{1988A&A...195L...1N, 2011ApJ...736L..26A, 2012ApJ...750..121G, 2015ApJS..217...32B, 2016MNRAS.457..320S}.\footnote{\citet{2015ApJS..217...32B} labelled many ES galaxies S0$^-$ sp/E5-E7. However, the (E1-E4)-looking galaxies can also be ES galaxies with either face-on or edge-on discs at interior radii.}  As noted in the Introduction of \citet{2024MNRAS.535..299G}, 
this population is similar, but not equal, to the `discy ellipticals' presented by \citet{1988A&A...195L...1N}, \citet{1989A&A...215..266N}, and \citet{1995A&A...293...20S}. The ES,b galaxies are compact and akin to bulges, while the ES,e subtype are extended and more like E galaxies \citep{Graham:Sahu:22b}.  
Residing near the top of the 
wet-major-merger-built S0 galaxy sequence, the ES,b galaxies may be relic mergers. 
 This would then rule them out as larger evolved counterparts of LRDs in the sense of evolution by simple gas accretion.  Red nuggets might, however, be morphed descendants of LRDs in the sense of a (major merger)-induced `punctuated equilibrium' event, i.e.\ collision, that transformed the primaeval discs into these more bulge-dominated galaxies at high redshift. 
 
Although
multiple stellar populations were not detected in the local ES,b galaxies Mrk~1216, NGC~1271, or NGC~1277
\citep{2015ApJ...808..183W, 2015MNRAS.452.1792Y} --- which might be odd given their bulge/disc nature if not for their old ages --- \citet{2011MNRAS.410.1223R} tentatively report different
populations in NGC~1332 (dust=y) 
and \citet{2019MNRAS.487.3776P}
measure differences within NGC~3115 (dust=Y), 
which is reported to have a spatially offset AGN due to a past merger \citep{2014ApJ...796L..13M}. 
The ES,b galaxy NGC~1277 became well known after it was thought to have an over-massive black hole giving an $M_{\rm bh}/M_{\rm \star,sph}$ ratio of 0.59 \citep{2012Natur.491..729V}. However, the black hole mass was overestimated by an order of magnitude and the spheroid mass underestimated by the same amount \citep{2016ApJ...819...43G}. As such, NGC~1277 is not considered an analogue of an LRD with a particularly high $M_{\rm bh}/M_{\rm \star,sph}$ ratio. 
The large, extended ES,e galaxies highlighted in the figures have similar $B/T$ ratios to the ES,b galaxies.  They are built by (perhaps lower redshift) mergers \citep[][and references therein]{2024MNRAS.535..299G}.  However, they have notably 
lower $M_{\rm bh}/M_{\rm \star,sph}$ ratios and lower stellar densities within
their spheroid's half-light radii \citep[][their Table~1]{Graham:Sahu:22b}.
%


\subsection{LRDs with AGN}
\label{Sec_LRDs_AGN}


Figure~\ref{Fig-M-M-sph} shows that NSCs and the dense inner components of UCD galaxies have stellar masses 2 to 3 orders of magnitude smaller than local galaxies with the same black hole mass.  Curiously, they roughly follow a distribution with a similar slope in the $M_{\rm bh}$-$M_{\rm \star,sph}$ diagram. Although new NSC and UCD galaxy data has been included (see Section~\ref{Sec_ucd_data}), the line shown on the left-hand side of Figure~\ref{Fig-M-M-sph} has been taken from \citet{2020MNRAS.492.3263G} and is essentially that first reported by \citet{2016IAUS..312..269G}. In a future paper, an updated analysis of this relation will be presented. The main objective here is to see how the LRDs compare.  In Figures~\ref{Fig-M-M-gal}--\ref{Fig-M-M-gal-plus}, the total (NSC $+$ envelope) stellar mass of the UCD galaxies is displayed.  
 
Many ideas exist as to how SMBHs established themselves early in
the Universe \citep[e.g.][and references therein]{2020ARA&A..58...27I}.  At first glance, the high (2023-2024 and 2025) SMBH
masses reported for the LRDs/AGN at high-$z$ may seem to favour heavy black hole seeds over light 
seeds, and one might even speculate that they could be primordial, contributing to
dark matter \citep[e.g.][]{1998PhLB..441...96A, 2002PhRvD..66f3505B, 2017JCAP...04..036D}.  
The direct collapse of
gas clouds \citep{1967SvA....11..233D, 1993ApJ...419..459U}
and self-gravitating pre-galactic gas disks
\citep{2006MNRAS.370..289B, 2006ApJ...652..902S} has been proposed for 
the production of the more massive seed black holes.  
At the 
same time, there can still be a contribution from light black hole seeds, including
those built from the cascading collision of stars in dense clusters at the
nuclei of haloes \citep{2002ApJ...576..899P, 2008ApJ...686..801O, 2021MNRAS.505.2186D}. In the
presence of substantial gas, dynamical friction will help feed these cluster stars
inward \citep{2008ApJ...686..801O, 2009ApJ...694..302D, 2011ApJ...740L..42D}.
This will contribute to the partial demise of the clusters and the
growth of the massive black holes. There is also 
near-Eddington \citep{2024NatAs...8..520W} and super-Eddington accretion
\citep{2005ApJ...633..624V, 2014Sci...345.1330A} onto these high-$z$ black holes, perhaps formed from
Pop III stars \citep{2016MNRAS.460.4122R, 2019MNRAS.483.3592B, 2024arXiv241201828T}.  Indeed,
\citet{2024arXiv240505333S} recently reported on LID-568, an SMBH at
$z\approx4$, accreting at 40 times the Eddington limit.  A combination of light and heavy black holes may have seeded the LRDs.

There may be clues at low-$z$ as to the evolution of LRDs.  The cE 
galaxies at $z\sim0$ are considered former low-mass ETGs stripped of many of
their stars \citep[e.g.][]{1988IAUS..126..603Z, 1993ASPC...48..608F, 2023Galax..12....1K}. 
At the same time, some rare isolated cEs may have simply never `grown up', unless they are all runaway systems that were stripped and then ejected from galaxy groups or clusters \citep{2015Sci...348..418C}. 
As illustrated in \citet[][their Figure~6]{Graham:Sahu:22b},
stripping of stars can produce galaxies with $M_{\rm bh}/M_{\rm \star,sph}\approx0.05$
(e.g. NGC~4486B around M87, or NGC~4342).  The more
extreme version of stripping, referred to as `threshing' \citep[e.g.][]{2003MNRAS.346L..11B,   
  2004ApJ...616L.107I, 2008MNRAS.385L..83C, 2013MNRAS.433.1997P}, is thought to produce UCD galaxies. 
  SDSS~J124155.33$+$114003.7 (M59cO)
  may represent a halfway point between cE and UCD galaxies \citep{2008MNRAS.385L..83C}.  Threshing can pare a galaxy back to $M_{\rm bh}/M_{\rm \star,sph}\sim 0.1$ (e.g.\ UCD1 around M60) or perhaps even $\sim$1 if just the NSC remains.

Given that the stripping process should preferentially remove the dark matter (thought to be dominant at larger, and thus less bound, radii)\footnote{Proponents of modified gravity offer an alternate view \citep[][their Section~5.1]{1983ApJ...270..384M, 2006JCAP...03..004M, 2010LRR....13....3D, 2012LRR....15...10F, 2020Univ....6..107D}.}, a key difference between local UCD galaxies formed via stripping and relic LRDs may be their dark matter fraction.
The lack of dark matter in UCD galaxies 
\citep[e.g.][]{2007A&A...463..119H, 2008MNRAS.390..906C, 2011MNRAS.412.1627C, 2011MNRAS.414L..70F} would seems to rule out the notion they are relic compact galaxies formed long ago in 
dark matter haloes, i.e.\ relic LRDs, as opposed to tidally-stripped galaxies. 
Furthermore, 
the {\em first} LRDs to form are perhaps unlikely to be today's UCD galaxies, as
these early LRDs probably resided in larger over-densities that eventually became
today's massive ETGs.  However, the LRDs that formed later in the Universe, in smaller
haloes, might be the ancestors of today's UCD galaxies, that is, they could be the nuclei of threshed low-mass disc galaxies that started life as LRDs.
One may also speculate that all of today's UCD galaxies should contain a massive black hole {\it if} all high-$z$ LRDs did. It is not, however, established that all local UCD galaxies contain a massive black hole, nor if all nucleated dwarf ETGs galaxies --- the likely pre-stripped progenitors of UCD galaxies --- do.  This state of affairs is an issue of spatial resolution and, therefore, probably does not yet provide useful constraints.


Based on the 2023-2024 masses for LRDs, they are not similar to NSCs for which
\citet{2009MNRAS.397.2148G} quantified their $M_{\rm bh}/M_{\rm \star}$ ratios.
Figure~\ref{Fig-M-M-gal} reveals that the LRDs tend to have smaller $M_{\rm bh}/M_{\rm \star}$ ratios, with some matching that of UCD galaxies.\footnote{The LRD with the exceptionally high mass is SDSS~J2236+0032 \citep{2023Natur.621...51D}. 
In Figure~\ref{Fig-M-M-gal}, it resides next to the ES,b galaxy NGC~1332 and the dust-rich S0 galaxy NGC~6861, suggesting it is more akin to a `red nugget' than an LRD, as does its half-light radius of 0.7$\pm$0.1~kpc in the F356W filter \citep{2023Natur.621...51D}.
}
Through the LRDs, we may be witnessing the reverse of a stellar stripping
process, in which the black hole mass is already in place or quickly forms before the bulk of a galaxy's stellar mass builds around it \citep[e.g.][their section 5.2]{2024ApJ...975..178K} and \citep{2024arXiv241204983T}.
However, with the revised black hole masses in the distant AGN/LRDs studied by \citet{2025arXiv250316595R}, 
Figure~\ref{Fig-M-M-gal-plus} reveals that they require more definitive stellar masses before firm conclusions can be reached.  
It is considered unlikely that the LRDs evolve along the steep $M_{\rm bh}$-$M_{\rm \star,gal}$ relations defined by local galaxies because if their current upper stellar mass estimates (shown in Figure~\ref{Fig-M-M-gal-plus}) are close to their true stellar masses, then their small half-light sizes (tens to a few hundred parsec) would imply a problematically high stellar density. 
LRDs likely have considerably lower stellar masses than the upper limits shown in Figure~\ref{Fig-M-M-gal-plus}, perhaps overlapping with the green peas or perhaps forming an extension towards the UCD galaxies and NSCs in the $M_{\rm bh}$-$M_{\rm \star,gal}$ diagram.

Finally, we are open to the possibility that whether UCD galaxies and some LRDs are structurally equivalent could be a case of `naïve realism' based on `the illusion of information adequacy' \citep{illusion}, stemming from their congruent location in the $M_{\rm bh}$--$M_{\rm \star,gal}$ diagram. 
%
Additional information may suggest that their overlap in SMBH and stellar masses is a mere coincidence.
In this regard, size information of LRDs and age estimation of UCD
galaxies \citep[e.g.][]{2008MNRAS.390..906C} should be valuable.
At least some UCD galaxies are measured to be reasonably old, such as the Sombrero galaxy's SUCD1 at 12.6$\pm$0.9 Gyr
\citep{2009MNRAS.394L..97H} and M60-UCD1 with a formal age of 14.5$\pm$0.5 Gyr
\citep{2014Natur.513..398S}, older than the Universe.  However,
the Virgo cluster's VUCD3 is reported to have an age of just 11~Gyr 
  with a 9.6-Gyr-old inner component, while M59cO has an age of 11.5~Gyr but
  with a blue inner component \citep{2008MNRAS.385L..83C} dated at just 5.5 Gyr old \citep{2017ApJ...839...72A}.  This
  may reflect the ability of star clusters to rejuvenate
  themselves, at least those residing at the bottom of a galaxy's
  gravitational potential well, or a late-time creation for some.  Detecting {\it elongated} LRD host galaxies with light profiles having S\'ersic indices $n \lesssim 1$ would suggest they are disc-like structures, whereas a distribution of somewhat spherical shapes would match that of local dETGs \citep[e.g.][]{1995AnA...298...63B}.  While single S\'ersic fits to galaxies with point-like AGN may yield small sizes, when the imaging data permits it, simultaneously fitting a point-source plus a S\'ersic model may enable more reliable information on the underlying host galaxy.

In passing, it is noted that the stellar population of LRDs 
(and red nuggets) would have initially been blue due to hot, massive stars, just as
the stars in the discs of today's low-mass disc-dominated S0 galaxies
would have been more\footnote{Due to low metallicity, dwarf S0 galaxies are at
the blue/green end of the `red sequence' \citep[][and references therein]{1959PASP...71..106B, 
  1961ApJS....5..233D, 1972MmRAS..77....1D, 
  2024MNRAS.531..230G}.} blue in the past when they were younger.
Ideally, future work will reveal further connections between LRDs and 
`green peas' \citep{2009MNRAS.399.1191C, 2024arXiv241208396L}, 
blue dwarf ETGs
\citep{2001ApJS..133..321C, 2003ApJ...593..312C, 2007ApJ...657L..85D, 2009ApJ...699..105C,
  2016MNRAS.457.1308M, 2019MNRAS.489.2830M}, 
  and luminous blue compact galaxies \citep{2003ApJ...586L..45G, 2005A&A...437..849S, 2006A&A...448..513B, 2011MNRAS.418.2350P}. 




\subsection{Additional AGN: toeing the line}
\label{Sec_toe}

As noted in the Introduction, the steep quadratic nature of the $M_{\rm bh}$--$M_{\rm \star,sph}$ relations for ETGs built from major mergers naturally place bright QSOs \citep[e.g.][]{2010ApJ...714..699W, 2014MNRAS.443.2077B, 2016ApJ...830...53W,   2017ApJ...845..138S, 2018ApJ...854...97D, 2016ApJ...816...37V} above the original near-linear relation.  
At the same time, the super-quadratic $M_{\rm bh}$--$M_{\rm \star,sph}$ (and
cubic $M_{\rm bh}$--$M_{\rm \star,gal}$) relation for S galaxies
naturally places faint Seyfert galaxies below the original near-linear relation. 
Bridging these extremes are the low-luminosity QSOs and regular AGN of
intermediate-luminosity \citep{2015ApJ...801..123W, 2017ApJ...850..108W,  
  2018PASJ...70...36I} that overlap with the original near-linear $M_{\rm  
  bh}$--$M_{\rm \star}$ relation. \citet{2021ApJ...914...36I} reported how the less luminous
QSOs beyond $z\sim6$ had $M_{\rm bh}/M_{\rm \star}$ ratios more in line with
the original near-linear $M_{\rm bh}$--$M_{\rm \star}$ relation.
Rather than talking in terms of differing and disjoint populations of AGN with
over- or under-massive black holes, it makes more sense to recognise that the
original proposition of a linear $M_{\rm bh}$--$M_{\rm \star}$ scaling
relation requires adjusting and that a steeper scaling relation unifies many galaxies.

Unfortunately, 
the lack of morphological information among AGN samples has hampered past understanding
of galaxy speciation.  
However, as briefly noted before, one slight mystery has now been resolved.
In Figure~\ref{Fig-M-M-sph}, and the $M_{\rm bh}$--$M_{\rm \star,sph}$ diagram presented in \citet{2015ApJ...798...54G}, some of the AGN were seen to reside to the right of the sample of galaxies with predominantly inactive black holes with directly measured masses.  This is evident at $M_{\rm bh}\approx10^7$ M$_\odot$. 
From the $M_{\rm bh}$--$M_{\rm \star,gal}$ diagram (Figure~\ref{Fig-M-M-gal}), it is apparent that these AGN are probably S galaxies.  It is likely that the published $B/T$ ratios for these AGN, which were typically greater than 0.5-0.6, were too high, thereby artificially shifting them to the right in the $M_{\rm bh}$--$M_{\rm \star,sph}$ diagram.  S galaxies tend to have $B/T<$ 0.1-0.2 \citep{2008MNRAS.388.1708G, 2019ApJ...873...85D}.  

Given the location of the AGN with $M_{\rm bh} \lesssim 10^6$ M$_\odot$ in Figure~\ref{Fig-M-M-sph} and Figure~\ref{Fig-M-M-gal}, they appear to be hosted by S galaxies and primaeval S0 galaxies. Again, the term primaeval is used to imply a first-generation galaxy type not altered by substantial accretion or major mergers.  These AGN are consistent with the steep morphology-specific $M_{\rm bh}$--$M_{\star}$ scaling relations, even though they were not used to define them.

A simplified variant of Figure~\ref{Fig-M-M-gal} is presented in
Figure~\ref{Fig-M-M-gal-plus}, such that the local sample of galaxies with
directly measured SMBH masses is now separated into just three types.
There are those previously identified as primaeval; they are the low-mass,
dust-poor S0 disc galaxies that tend to have an old, metal-poor stellar
population.
Second are the disc galaxies with a spiral pattern, which tend to have
ongoing star formation.\footnote{Arguably, gas-stripped and faded S galaxies, which are now S0 galaxies, belong to this category.}
Then there are the galaxies built from major mergers, such as the (often
dust-rich) S0 galaxies built from a wet major merger, the E galaxies built
from a dry major merger, and the BCG typically built from more than one major merger.
This subdivision offers an alternative view of how the local AGN mesh with, rather than deviate from, galaxies with varying formation histories. 

%

Figure~\ref{Fig-M-M-gal-plus} also displays new samples of AGN. 
The AGN with estimated black hole masses from \citet{2015ApJ...813...82R} and
\citet{2018ApJ...863....1C} are included, as are 
the AGN at $z\gtrsim6$ from \citet{2021ApJ...914...36I}.  However, only dynamical, rather than
stellar, galaxy masses are published for this final sample.
Therefore, a (not overly prominent) small circle is used to
show those systems in Figure~\ref{Fig-M-M-gal-plus}.  While \citet[][their
  Figure~13]{2021ApJ...914...36I} reported that many of these systems have
`overmassive' black holes (by up to a factor of $\sim$10) relative to the old
near-linear $M_{\rm bh}$-$M_{\rm \star}$ relation, the bulk of them are not 
outliers relative to the steeper galaxy-morphology-dependent $M_{\rm  
  bh}$-$M_{\rm \star}$ relation for merger-built S0 galaxies, a point made by
\citet{Graham-S0}, see also \citet{Graham:Sahu:22a}. Just 3 to 5 of these AGN from \citet{2021ApJ...914...36I} 
have $M_{\rm bh}/M_{\rm dyn}$ ratios that are a factor of (only) 2 to 3 times higher than
the distribution seen for local systems with directly measured SMBH
masses.
This is reconcilable with the sample selection bias of the most luminous QSOs at high-$z$ and the 
greater inaccuracy of indirect measures of black hole mass in AGN. 
The subset of luminous $z\gtrsim6$ AGN from \citet{2021ApJ...914...36I} shown in
Figure~\ref{Fig-M-M-gal-plus} is likely to be wet-merger-built S0 galaxies. 

At $M_{\rm \star,gal} > 10^{10}$ M$_\odot$, in Figure~\ref{Fig-M-M-gal-plus},  
the overlap of low-$z$ AGN with the $z\approx0$ sample of SMBHs with directly
measured masses reveals that these AGN are predominantly S galaxies or merger-built S0 galaxies.  These AGN are not an offset population from ordinary galaxies with predominantly inactive SMBHs.  This information is crucial if we are to connect the evolutionary path of high-$z$ AGN/LRDs with other active and inactive galaxies.

Knowledge of galaxy-morphology-specific black hole scaling relations 
enables an improved means for deriving the virial factor(s) for converting AGN virial products into black hole masses \citep{1993PASP..105..247P, 2004ApJ...615..645O, 2009ApJ...694L.166B}. 
To date, these conversion factors have been obtained with little attention to galaxy morphology.
A better approach will involve matching AGN virial products with directly measured black hole masses from galaxies {\em of the same morphological type} (Graham et al.\ 2025, in preparation). 
It can already be seen in the data of \citet[][their Figure~5]{2018ApJ...864..146B} that the reverberation-mapped AGN follow a steep $M_{\rm bh}$--$M_{\rm \star,sph}$ distribution well-matched by the super-quadratic relation quantified by \citet[][their Figure~3]{2013ApJ...768...76S} for galaxies without depleted stellar cores, i.e.\ those not built from a dry major merger.  Similarly, the AGN data of \citet[][their Figure~6]{2018ApJ...864..146B} display a steep trend similar to that of \citet{2016ApJ...817...21S} for S galaxies.   
This yet-to-be-applied approach at determining the virial factor(s) will avoid a bias such that the reverberation-mapped AGN sample may be skewed toward a distribution of galaxy type, such as S and dust-rich S0 galaxies, that differs from the bulk of the sample with directly measured black hole masses to which it is compared. 
Similarly, AGN sample selection of certain galaxy types and not others, for example, S and/or dust-rich S0 but not primaeval dust-poor S0 galaxies or `red and dead' E galaxies, may also explain the tight AGN relations reported by \citet{2021ApJ...921...36B}. 
The location of the LRDs and other AGN in Figures~\ref{Fig-M-M-sph}--\ref{Fig-M-M-gal-plus} can be refined once improved virial factors, used to calibrate secondary relations, are established through application of the galaxy-morphology-dependent scaling relations \citep{Graham-triangal}.  Better constraints on the stellar masses of the LRDs is also needed to decipher their evolutionary trajectory in the $M_{\rm bh}$-$M_{\star}$ diagrams.


\section{Summary}\label{Sec_Sum}

This paper added high-$z$ AGN/LRDs and low-$z$ AGN to $M_{\rm bh}$-$M_{\star}$ diagrams that, for the first time, included local compact stellar systems (UCD galaxies and NSCs) in addition to larger galaxies with directly measured black hole masses. Our diagrams also included an expanded recognition of local galaxy morphologies and galaxy-morphology-specific $M_{\rm bh}$-$M_{\star}$ relations rather than a single near-linear relation for low-$z$ galaxies with AGN and a separate single near-linear relation for $z\approx0$ inactive galaxies.
Unlike previous $M_{\rm bh}$-$M_{\star}$ relations based on various fractions of ETGs and LTGs, or all of the different ETGs combined, these morphology-specific relations avoid sample selection bias from mixing different galaxy types that follow distinct $M_{\rm bh}$-$M_{\star}$ relations. 

The 2023-2024 LRD data is seen to span the $M_{\rm bh}$-$M_{\rm \star,gal}$ diagram from UCD galaxies to previously recognised primaeval S0 galaxies.  With only upper limits on the recent 2025 stellar masses of LRDs \citep{2025arXiv250316595R}, they remain consistent with local NSCs and UCD galaxies, low-$z$ green peas, and primaeval local S0 galaxies.  An improved knowledge of the LRD stellar masses is needed.

We demonstrated that several samples of low-$z$ AGN, including candidate intermediate-mass black holes \citep{2015ApJ...798...54G, 2018ApJ...863....1C}, broadly overlap with the steep, non-linear, galaxy-morphology-dependent $M_{\rm bh}$-$M_{\star}$ relations defined by predominantly inactive galaxies. 
This suggests that using galaxy samples with similar morphology should be worthwhile for re-determining the virial factor used to estimate black hole masses in LRDs and AGN more broadly.  Adjustments to high~$z$ QSO and LRD black hole masses will impact expectations for black hole seed masses, early accretion rates, and connecting the past with today.


\section*{Acknowledgements}

This paper is dedicated to Alexander Bolton Graham (1933-2024), 
who adopted AWG many years ago and patiently listened (as a frequent hospital patient in 2020-2024) to unpublished developments in galaxy/black hole research.


Igor Chilingarian’s research is supported by the SAO Telescope Data Center. He also acknowledges support from the NASA ADAP-22-0102 and HST-GO-16739 grants. DDN's work is partially supported by a grant from the Simons Foundation to IFIRSE, ICISE (916424, N.H.). 
This research has used the NASA/IPAC Extragalactic Database (NED) and
the SAO/NASA Astrophysics Data System (ADS) bibliographic services.

\bibliographystyle{mnras}
\bibliography{LRD_UCD}

\end{document}